\documentclass[a4paper,fleqn]{cas-dc}

\usepackage[authoryear,longnamesfirst]{natbib}
\usepackage[switch]{lineno}

\def\tsc#1{\csdef{#1}{\textsc{\lowercase{#1}}\xspace}}
\tsc{WGM}
\tsc{QE}

\begin{document}
\let\WriteBookmarks\relax
\def\floatpagepagefraction{1}
\def\textpagefraction{.001}

\shorttitle{Predicting the First Martian Meteors}    

\shortauthors{M. Vovk et~al.}  

\title [mode = title]{From Earth Meteors to Mars: Predicting Where to See the First Martian Meteors}  





\author[1,2]{Maximilian Vovk}[orcid=0000-0002-9738-771X]
\cormark[1]
\ead{mvovk@uwo.ca}
\credit{Conceptualization, Methodology, Software, Data curation, Writing - original draft}

\affiliation[1]{organization={Department of Physics and Astronomy, University of Western Ontario},
            addressline={1151 Richmond Street}, 
            city={London},
            postcode={N6A 3K7}, 
            state={Ontario},
            country={Canada}}

\affiliation[2]{organization={Institute for Earth and Space Exploration, University of Western Ontario},
            addressline={Perth Drive}, 
            city={London},
            postcode={N6A 5B8}, 
            state={Ontario},
            country={Canada}}

\author[1,2]{Peter G. Brown}[orcid=0000-0001-6130-7039]
\ead{pbrown@uwo.ca}
\credit{Conceptualization, Methodology, Writing - review \& editing}

\author[1,2]{Denis Vida}[orcid=0000-0003-4166-8704]
\ead{dvida@uwo.ca}
\credit{Conceptualization, Methodology, Software, Writing - review \& editing}

\cortext[1]{Corresponding author}


%



\begin{abstract}
Predictions of optical meteors at Mars have largely relied on classical single-body ablation models, despite high-resolution terrestrial observations showing that mm-sized meteoroids frequently fragment. This study quantifies how fragmentation alters the predicted brightness and peak-luminosity altitudes of sporadic mm-sized meteoroids in the Martian atmosphere and evaluates the single-body approximation as a reference baseline. Physical properties were inferred for 144 sporadic meteoroids observed on Earth using dynamic nested sampling with the erosion--fragmentation model. The resulting best-fit meteoroids were then re-simulated under Martian atmospheric conditions to generate predicted light curves. Because the physical trigger of fragmentation onset remains uncertain, three hypotheses were tested based on atmospheric mass density, dynamic pressure, and total accumulated heat. The results were also compared with predictions from a single-body ablation model. The data-driven simulations predict peak absolute magnitudes of \(M_{\rm peak}\sim 2\)--7 for Martian meteors spanning diameters of (0.4)--(10)~mm and entry speeds of 10--56~km\,s\(^{-1}\). We find most events are luminous between $\sim$ 55 and 110~km heights. Relative to the single-body ablation baseline, the fragmentation-based predictions are brighter by $\sim$ 0.8~mag at peak brightness and concentrate luminosity within a narrower vertical range ($\sim$ 17~km instead of 36~km). The modelling accounting for fragmentation also produce shorter luminous trails ($\sim$ 20~km instead of 45~km). The resulting altitude--brightness maps provide observation-ready guidance for future Mars missions, while the fragmentation-based framework supports the interpretation of meteoroid-related ionospheric metal layers and improvements to Mars meteoroid-environment models.
\end{abstract}

\begin{graphicalabstract}
\includegraphics[width=\columnwidth]{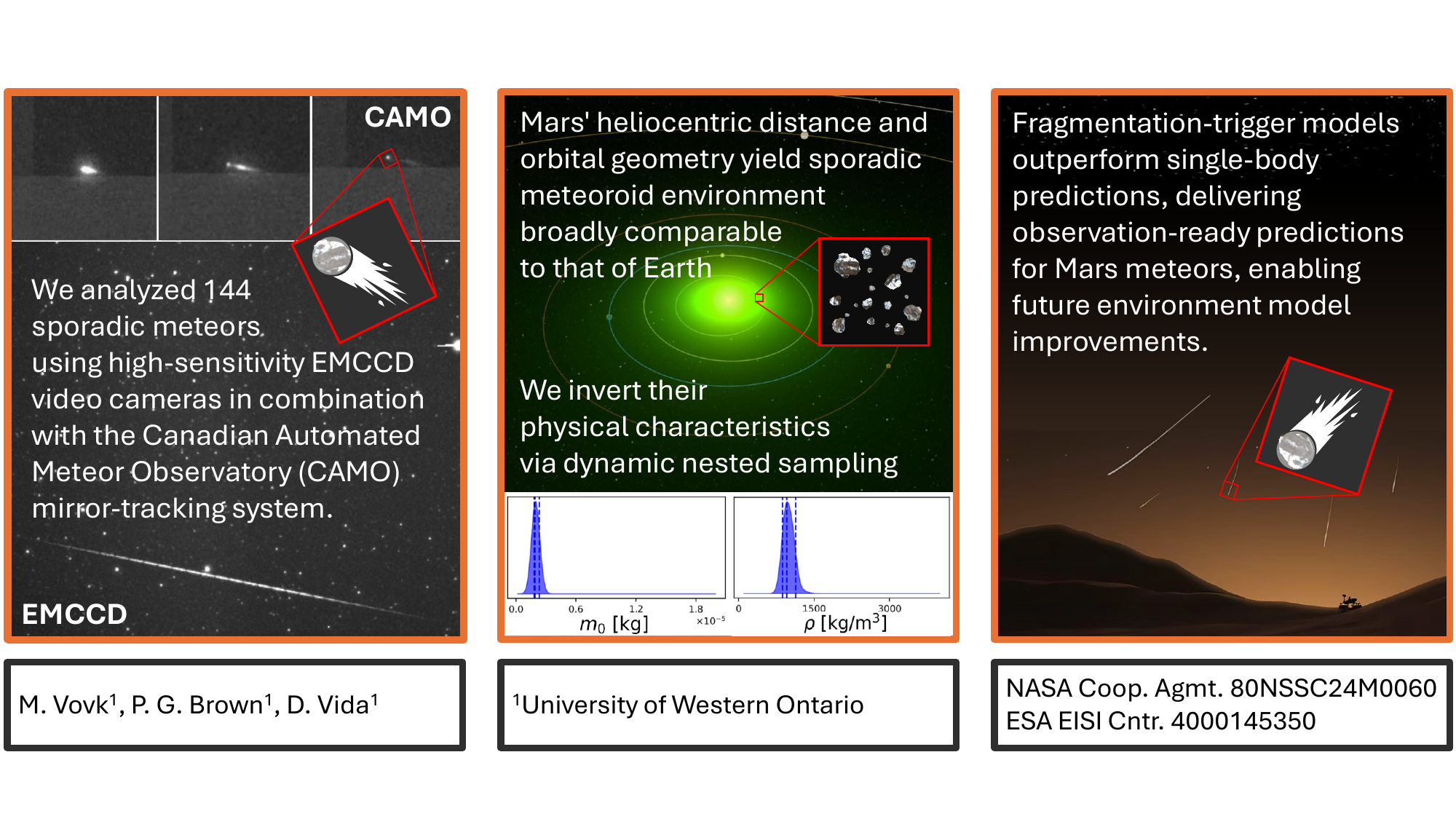}
\end{graphicalabstract}

\begin{highlights}
    \item Earth meteors predict detectability for future Mars missions.
    \item Three fragmentation triggers assumed to produce distinct Martian meteors.
    \item Fragmentation makes Martian meteors shorter than single-body predictions.
    \item mm-sized Mars meteoroids peak at \(\sim 2\) to 7 mag between 55 and 110 km.
\end{highlights}


\begin{keywords}
 Meteors \sep Meteoroids \sep Mars \sep Planets and satellites: atmospheres
\end{keywords}

\maketitle

\section{Introduction}

Planetary atmospheres throughout the Solar System are continually modified by the infall of interplanetary dust and meteoroids. On Earth, this influx not only seeds the upper atmosphere with metals and smoke particles, but also produces optical meteor phenomena that have been extensively characterised by radar and optical observations \citep{ceplecha1998meteorTheory}. Extending such studies beyond Earth is essential for understanding how meteoroid influx shapes the upper atmospheres, ionospheres, and cloud microphysics of other terrestrial planets \citep[e.g.][]{Whalley2010}. 

In addition, physically constrained meteor and ablation predictions and measurements provide one of the best pathways to improving meteoroid environment models. Major engineering-focused models, such as NASA’s Meteoroid Environment Model (MEM) and ESA’s Interplanetary Meteoroid Environment Model (IMEM), are largely unconstrained for larger particle sizes in regions of the solar system beyond Earth. Linking meteoroid populations to observable atmospheric signatures on other planets is crucial for the validation of these models in other areas of the solar system. Additionally, such modeling and measurements are key for estimating meteoroid mass deposition profiles and models of upper atmospheric chemistry \citep{Plane_Flynn2018}.

Mars' heliocentric distance and orbital geometry yield a meteoroid environment broadly comparable to that of Earth, with a daily influx of meteoroid material from the zodiacal cloud estimated at $2.1\pm1.0$~t \citep{Carrillo-Sanchez2019a}. The Martian atmosphere is overwhelmingly composed of CO$_2$ (96\%), and its surface pressure of $\sim 0.00628$~atm is comparable to the pressure at $\sim 120$~km altitude in the terrestrial atmosphere \citep{lissauer2019fundamentalplanetary}. Given its lower atmospheric total column mass density, meteoroids with diameters of $\sim 0.60$–$1.2$mm can partially or fully survive atmospheric entry and potentially reach the surface unmelted \citep{Flynn1990mars}. Early modeling suggests that, for a given apparent magnitude, Mars would exhibit roughly half as many meteors as Earth, with most Martian meteors occurring between 50 and 90~km altitude \citep{Adolfsson1996} and reaching peak brightness between 75 and 85~km \citep{molina2003meteoric}.

Despite the absence, as yet, of a confirmed optical meteor observation at Mars \citep{Domokos2007, Christou_Vaubaillon2019}, several atmospheric and ionospheric tracers provide indirect proof of meteoroid interactions. On Earth, noctilucent clouds nucleate on meteoric smoke particles; analogous high-altitude CO$_2$ ice clouds have been identified on Mars from both orbiter and rover observations. Measurements from Mars Express and MAVEN have revealed ionised metallic layers at altitudes of $\sim$80–90~km, coincident with the expected ablation heights of meteoroids, and suggest a possible link between the spatial distribution of dry-ice clouds and meteor activity \citep{Hartwick2019_noctudileMars}. 

On Mars, Mg and Fe ions released during ablation predominantly undergo charge exchange with atmospheric ions, forming persistent metal layers \citep{Molina2008MetallayersMARSVENUSTITAN}. Such meteoroid-associated ionospheric layers have been detected by multiple missions, including Mariner~IV, Mars~4 and~5, Mars Express, and MAVEN \citep{Carrillo-Sanchez2019a}, while Mars Express has specifically confirmed a sporadic metal layer of Mg and Fe ions between 65 and 110~km exhibiting substantial temporal variability \citep{Molina2008MetallayersMARSVENUSTITAN}. These variations have been attributed to either meteor showers from cometary encounters or to impact ionisation by penetrating solar-wind ions, as in meteoric proton aurorae \citep{Crismani2022meteoiricProtonMars}.

However, current interpretations of these tracers rely almost exclusively on single-body ablation models originally developed for Earth, in which meteoroids are treated as intact objects that lose mass smoothly via vaporization as they decelerate \citep{opik1958physics}. Contemporary high-resolution video observations at Earth of meteors show that this assumption is rarely valid: nearly 90\% of faint meteors exhibit clear evidence of fragmentation, while the light curves of the rest cannot be explained by single-body ablation \citep{Subasinghe2016}. At Mars, meteoroids of order a few millimetres in diameter are expected to undergo substantial ablation \citep{Pabari2023_venusMars} and can contribute efficiently to optical meteor production, precisely the size range where fragmentation is found to be ubiquitous at Earth. Applying single-body ablation to such particles at Mars is therefore likely to bias predictions of where energy and mass are deposited, and of the altitudes at which their luminosity and metal layers peak.

Fragmentation models span a range of physical scales and levels of structural complexity. For fireballs and asteroid-scale bodies (m-sizes and larger), the fragment-cloud model combines progressive discrete fragmentation with the release of debris clouds to reproduce atmospheric energy-deposition profiles and investigate pre-entry structure \citep{Wheeler2018}. More recently, sintered discrete-element modelling has been used to explicitly represent irregular geometry, internal bonds, structural failure, and the motion of individual fragments during the disintegration of the Morávka meteoroid \citep{Li2026}. These approaches provide detailed descriptions of the fragmentation of comparatively large bodies, whereas the semi-empirical erosion--fragmentation model from \citet{borovivcka2007atmospheric} used here is intended to reproduce the gradual release of grains from millimetre-sized meteoroids constrained by high-resolution photometric and dynamical observations.

The aim of this study is to quantify how observationally-driven fragmentation behaviour found for meteors at Earth changes the predicted brightness and ablation profiles of sporadic meteors in the Martian atmosphere, relative to traditional single-body models. We focus specifically on millimetre-sized meteoroids, for which fragmentation is commonly observed at Earth and use the erosion--fragmentation model which is well suited to reproducing high-resolution photometric and dynamical measurements. By doing so, we seek to provide a more physically grounded framework for interpreting existing metal-layer observations and for planning future meteor-detection experiments at Mars.

To this end, we employ the erosion–fragmentation model of \citet{borovivcka2007atmospheric}, which has been successfully used to reproduce high-resolution optical observations of mm-sized meteors at Earth. We first use this model to infer the physical properties of a sample of 144 sporadic meteors in the terrestrial atmosphere, constraining physical parameters such as mass, bulk density, and fragmentation behaviour. We then project these meteoroids to Mars, considering the different atmospheric environment and entry speed at Mars, and re-simulate their atmospheric entry in the Martian environment using the same inferred meteoroid physical parameters, and compute the resulting light curves profiles. By comparing the fragmentation-model results with those obtained from a standard single-body ablation model for the same meteoroids, we assess how fragmentation modifies the altitude at which mass and energy are deposited and construct luminosity maps that indicate at which heights such mm-sized sporadic meteoroids would be observable on Mars.

\section{Methods}\label{sec:mars_method}





We analyzed 144 sporadic meteors using high-sensitivity Electron Multiplying Charge-Coupled Device (EMCCD) video cameras in combination with the Canadian Automated Meteor Observatory (CAMO) mirror-tracking system operating at two sites (Elginfield: 43.19279$^{\circ}$~N, 81.31565$^{\circ}$~W; Tavistock: 43.26420$^{\circ}$~N, 80.77209$^{\circ}$~W). This video system is operated at 32 frames per second and has a stellar limiting magnitude of about $+8$. The cameras provided time-resolved measurements of trajectory, velocity, deceleration, and light curves for meteors of limiting peak brightness of roughly +7 that were required for physical entry modelling. We provide a brief summary of the data and reduction process. 

To model each meteor, we followed the approach of \citet{vovk2026inferring}. We used the dynamic nested sampling (DNS) framework implemented in \texttt{dynesty} \citep{speagle2020dynesty} to infer meteoroid physical parameters within the erosion--fragmentation model of \citet{borovivcka2007atmospheric}, adopting the luminous efficiency of \citet{vida2024first}. The ablation model included two erosion stages to allow for changes in fragmentation behaviour during flight, where different 10 - 500 micrometre grains are released. All simulations were initiated at an entry altitude of 180~km in the terrestrial atmosphere. The likelihood function was constructed to balance the fit between the measured light curve and the observed deceleration profile, accounting for measurements noise in both. Additional details of the DNS implementation are provided in Appendix~\ref{sec:AppendixDynami_nested}.

To enable a direct comparison of our results and predictions with the modelling approach most commonly adopted for meteors on other planets, we also fitted each event with a single-body ablation model. We ran DNS using a simplified forward model based on the classical single-body formulation of \citet{opik1958physics}, in which the meteoroid is treated as an intact body that loses mass smoothly through ablation alone (i.e. atom-by-atom vaporization) and produces luminosity only from the main body. This formulation is attractive for planetary applications because it involves relatively few free parameters, minimal tuning and the single body solutions are analytical solvable \citep[e.g.][]{Molina2008MetallayersMARSVENUSTITAN, Plane_Flynn2018}. However, single-body ablation models are widely understood to not reproduce the full set of observed dynamics and light-curve structure for mm-sized meteoroids \citep{popova2019modelling} as these experience fragmentation \citep{Subasinghe2016} and generally do not agree with the measured photometric and dynamical constraints. We therefore treated the single-body solutions strictly as a reference baseline to represent previous studies.


\begin{figure}
\centering
\includegraphics[width=\columnwidth]{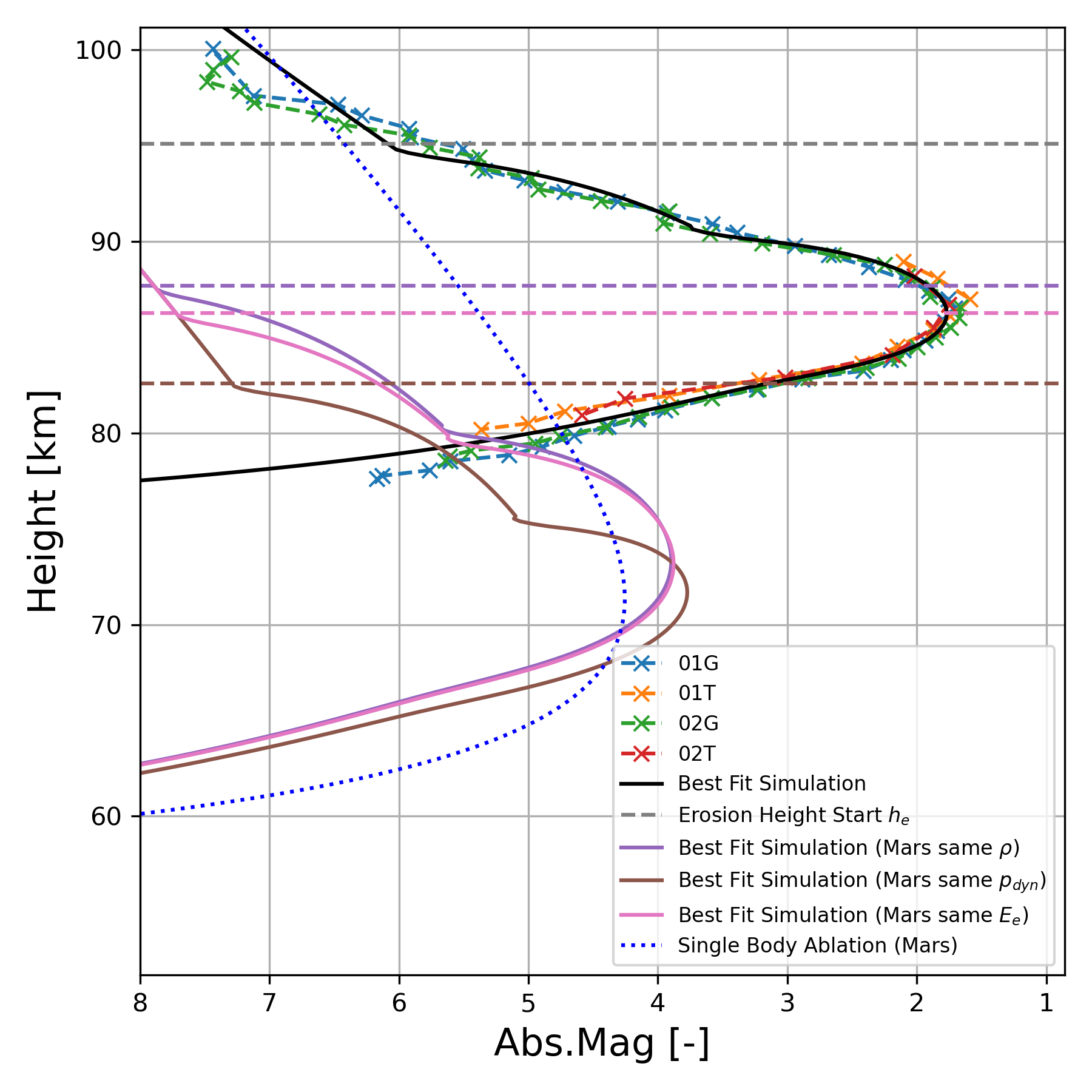}
\caption{Example of an observed Earth meteor light curve and simulated Mars light curve for 20220722\_050117 meteor (speed on Earth 20.2 km/s, Mars 15.3 km/s). The observed EMCCD light curves from two stations (01G, 02G) and CAMO (01T, 02T) are shown together with the best-fit erosion--fragmentation model on Earth (black; maximum-likelihood DNS realization). The Mars predictions for the same meteoroid, using identical intrinsic physical parameters but different mappings of the erosion onset condition and a real Martian atmosphere, are shown in shades of pink for the $\rho$-, $p_{\rm dyn}$-, and $E_e$-trigger cases. The corresponding Mars single-body ablation model prediction is shown as a dotted blue line.}
\label{img:mars_lightcurve}
\end{figure}

For each observed meteor, we selected the maximum-likelihood realisation from the posterior of the Dynesty output runs as the representative best-fit solution. This approach yields a single set of inferred physical parameters (e.g., mass, bulk density, ablation and erosion coefficient, erosion height) per meteor that best reproduces the Earth-based observations. These best-fit meteoroids were then propagated to Mars and re-simulated under Martian atmospheric conditions using the same inferred intrinsic physical parameters. This required adopting an appropriate Mars atmospheric density model as detailed in Appendix \ref{sec:MarsAtmosphere} and specifying the corresponding entry conditions at Mars (in particular the entry speed), as described in Appendix \ref{sec:speed_mars}.


The single-body model can be transferred to a different planet directly by changing the atmospheric density profile and entry conditions, because it contains no explicit fragmentation onset criterion. In contrast, the erosion--fragmentation model requires a trigger that determines when erosion begins at height $h_e$ and, in our two-stage formulation, when the second erosion stage $h_{e2}$ is activated. There is an ongoing debate in the field about whether fragmentation is triggered mechanically or thermally \citep{popova2019modelling, henych2026geminids}, we therefore considered three physically motivated hypotheses for the onset of erosion:

\begin{enumerate}
\item A simple constant atmospheric density trigger, in which erosion begins when the local atmospheric density matches the value at the inferred onset on Earth ($\rho$-trigger).
\item A constant dynamic pressure trigger, in which erosion begins when the dynamic pressure ($p_{\rm dyn} = \rho_{\rm atm} v^2$) matches the value at the inferred onset on Earth ($p_{\rm dyn}$-trigger). The assumption is that the onset of erosion is mechanical and grains start to be detached from the surface of the meteoroid due to mechanical forces.
\item A constant accumulated heat trigger, in which erosion begins when the cumulative energy received up to that point matches the value at the inferred onset on Earth ($E_e$-trigger). The assumption is that this energy represents the total received heat and the grains start to be detached from the surface of the meteoroid due to thermal degradation of a ``glue'' that holds the grains together. The computation is explained in detail in Appendix \ref{sec:erosion_trigger_mapping}.

\end{enumerate}


\begin{figure}
\centering
\includegraphics[width=\columnwidth]{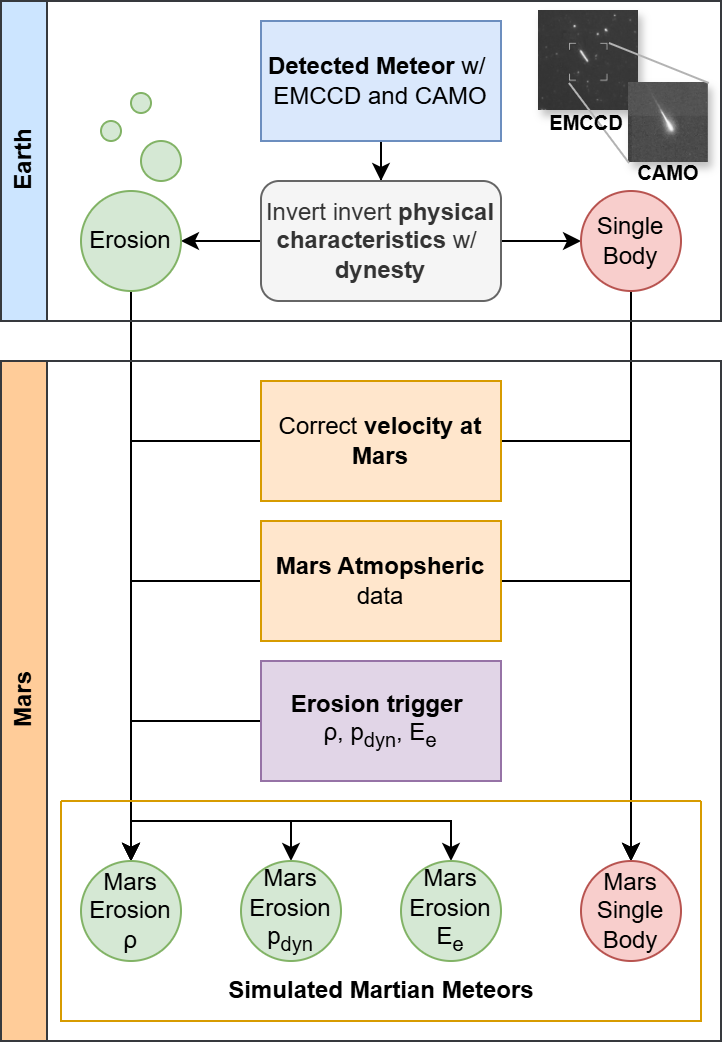}
\caption{Flowchart of the modelling workflow used to project Earth-observed meteors to Mars. Meteoroid physical properties are first inferred from Earth observations using dynamic nested sampling, then re-simulated in the Martian atmosphere for a corrected velocity at Mars and using either erosion--fragmentation trigger mappings or a single-body ablation model.}
\label{img:mars_flowchart}
\end{figure}

Consequently, for each analyzed meteoroid we obtained four Mars light-curve predictions: three erosion--fragmentation light curves (for the $\rho$, $p_{\rm dyn}$, and $E_e$ triggers) and one single-body ablation light curve. An example is shown in Figure~\ref{img:mars_lightcurve} and the complete flowchart is shown in Figure~\ref{img:mars_flowchart}.

\section{Results}






\begin{figure*}
\centering
\includegraphics[width=\textwidth]{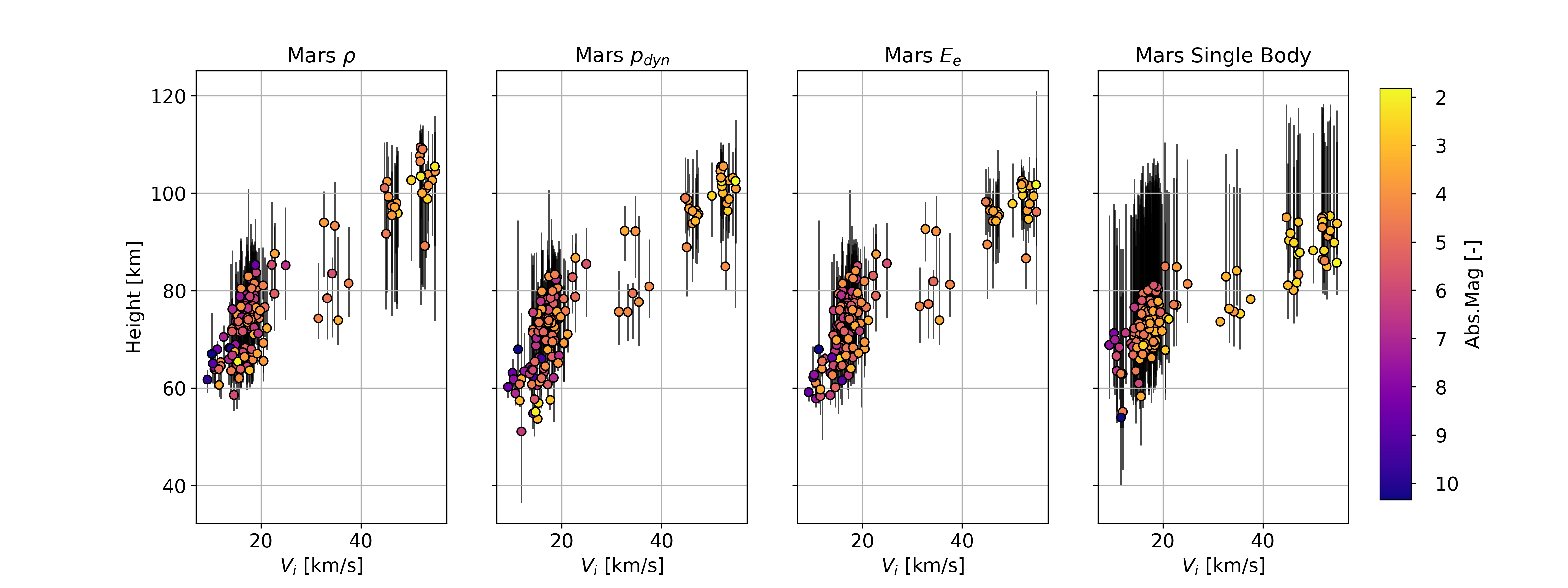}
\caption{Predicted peak brightness for meteors observed at Earth whose physical properties have been inverted in the Martian atmosphere as a function of Martian entry speed and altitude. Each panel shows, for one modelling assumption, the absolute magnitude at the Martian meteor predicted light-curve peak (colour scale) plotted against the meteoroid entry speed at Mars and the altitude of peak brightness, while the black line shows the altitude for a detection threshold of $\Delta M = 2.5$. The four panels correspond to the erosion--fragmentation model with three mappings of the erosion onset condition ($\rho$-trigger, $p_{\rm dyn}$-trigger, and $E_e$-trigger) and to the single-body ablation model. All magnitudes are in the GAIA G-bandpass.}
\label{fig:all_plts_bright}
\end{figure*}

The meteoroids inferred from the Earth observations span diameters of $\sim 0.4$--$10$~mm and masses from $1.6\times10^{-6}$ to $2\times10^{-4}$~kg. Propagated to Mars, they have expected entry speeds of 10--56~km\,s$^{-1}$. For each simulated Martian light curve, to define the trail length we adopted a detection threshold of $\Delta M = 2.5$ from the peak \citep{fleming1993light}, and extracted the initial height $h_i$, final height $h_f$, peak height $h_{\rm peak}$, peak absolute magnitude $M_{\rm peak}$, and trail length $L$. The computation of these parameters is described in Appendix \ref{sec:detection_threshold}.

Because the atmospheric density profile and entry speed differ substantially between Earth and Mars, these mappings generally imply different erosion onset altitudes and hence different predicted light curves. 


\subsection{Differences in modeled meteors at Mars}

Figure~\ref{fig:all_plts_bright} shows the predicted Martian peak brightness as a function of entry speed and peak altitude. For meteoroids detectable on Earth, the Martian peaks typically fall in the range $M_{\rm peak}\sim 2$--7 and $h_{\rm peak}\sim 50$--110~km. The fragmentation cases produce a broader peak-height distribution and a stronger speed dependence than the single-body baseline, which remains confined to a narrower range of $\sim 55$--95$~km$.

The single-body solution also predicts a larger luminous height range. Its median initial height is $100.5$~km, compared with $84.4$--$85.9$~km for the fragmentation cases, while the median final heights are $64.9$~km for the single-body case and $67.6$--$68.6$~km for the fragmentation cases. As a result, the single-body model gives a median luminous height range of $35.6$~km, compared with only $16.8$--$17.3$~km for the fragmentation cases, and a median trail length of $45.3$~km instead of $19.5$--$21.8$~km, much closer to the median value on Earth of $19.2$~km. Among the fragmentation solutions, the $\rho$ trigger starts slightly higher, while the $p_{\rm dyn}$ and $E_e$ cases remain very similar because they depend on both atmospheric density and the entry conditions.

\begin{figure}
\centering
\includegraphics[width=\columnwidth]{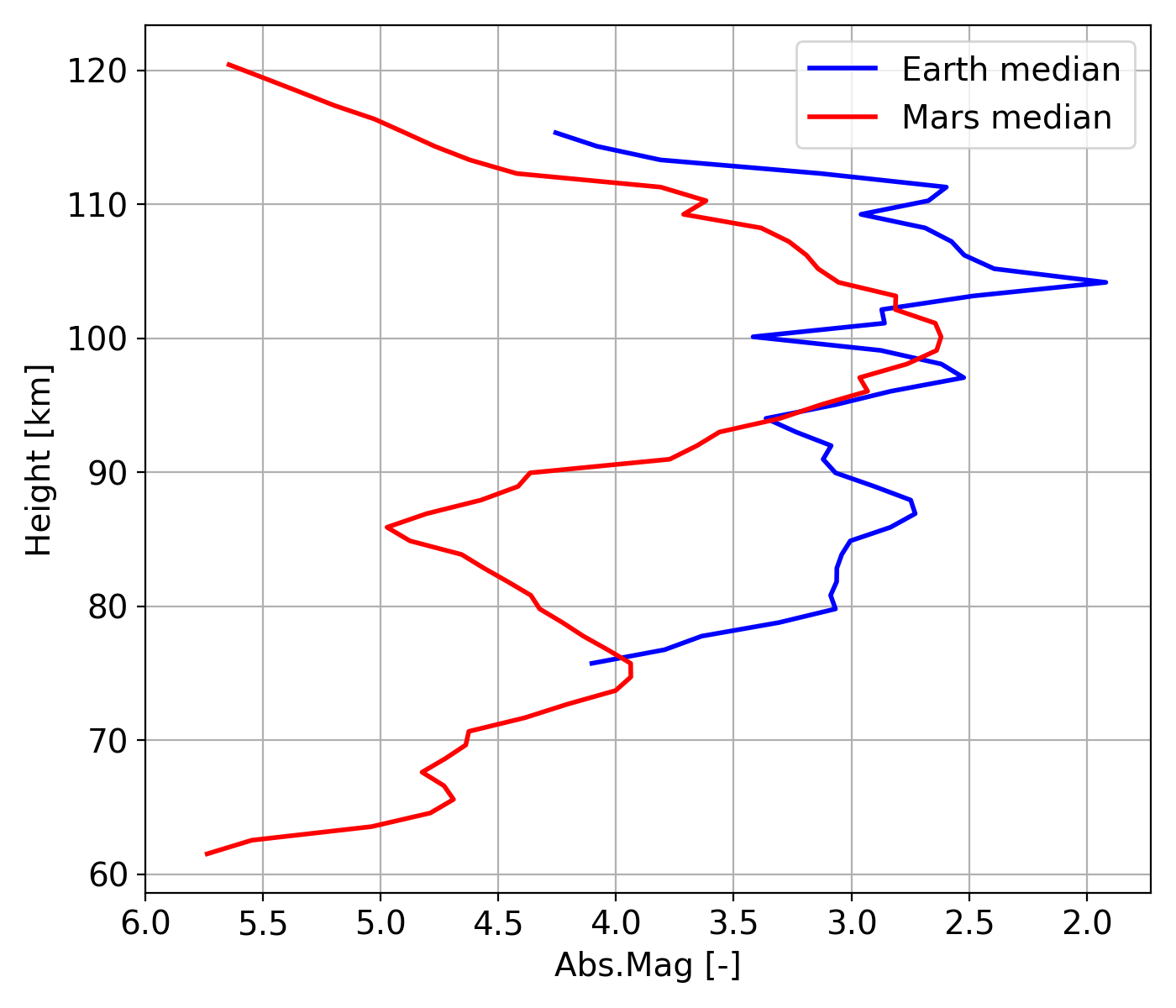}
\caption{Median absolute-magnitude profiles as a function of height for the Earth observations and the corresponding Mars predictions (erosion--fragmentation model). The two-peak structure reflects the presence of two dominant speed populations, which deposit their luminosity over different altitude ranges.}
\label{img:mars_distribution_abs_mag}
\end{figure}

\subsection{Brightness change relative to Earth}

Applying the same $\Delta M = 2.5$ detection threshold to the Earth observations, meteors observed at Mars allowing for fragmentation are systematically fainter. The median absolute magnitude dimming is about $2$~mag for slow entries (10--20~km\,s$^{-1}$) at around 80 km, about $1$~mag at intermediate speeds, and only $\sim 0.2$--0.5~mag for the fastest entries (40--50~km\,s$^{-1}$) at around 100 km compared to Earth. Thus, slow meteoroids reach only $M_{\rm peak}\sim 6$--7 on Mars, compared with $\sim 2$--3 on Earth, whereas fast meteoroids can reach equivalent peak magnitudes on both planets, a result  similarly found by \cite{Adolfsson1996}.

Figure~\ref{img:mars_distribution_abs_mag} shows the median magnitude as a function of height. The fast population peaks at similar brightness but slightly deeper on Mars, whereas the slower population shifts to lower altitudes and is fainter by about $1$~mag, mainly because of the different entry velocities at Mars and the lower atmospheric densities relative to Earth at the relevant ablation heights (see Appendix~\ref{sec:MarsAtmosphere}).

\section{Discussion and Conclusion}\label{sec:disc_conc}





This study quantified how realistic fragmentation changes the predicted brightness and peak-luminosity altitudes of mm-sized sporadic meteoroids at Mars relative to the single-body ablation assumption commonly adopted in previous Mars meteor predictions. Best-fit physical parameters were inferred for 144 Earth-observed sporadic meteors using dynamic nested sampling with the erosion--fragmentation model of \citet{borovivcka2007atmospheric} and the luminous-efficiency model of \citet{vida2024first}. These meteoroids were subsequently propagated to Mars and re-simulated under Martian atmospheric conditions. Because erosion onset cannot be constrained directly at Mars, three physically motivated trigger mappings were considered: atmospheric density \(\rho\), dynamic pressure \(p_{\rm dyn}\), and accumulated energy \(E_e\). A classical single-body model \citep{opik1958physics} was also evaluated as a reference baseline for comparison with previous studies. A meteor was considered detectable when it became brighter than the faintest portion of the corresponding event observed on Earth.

The principal result is that fragmentation-inclusive modelling predicts a broader and more speed-dependent distribution of peak-brightness altitudes than single-body ablation (Fig.~\ref{fig:all_plts_bright}). Another significant difference is in the predicted trail length: the fragmentation cases produce luminous vertical extents of 16.8--17.3~km, compared with 35.6~km for the Martian single-body solutions. Median trail lengths decrease from 45.3~km in the single-body case to 19.5--21.8~km in the fragmentation cases, approaching the typical value observed at Earth (19.2~km). The median peak brightness also changes from \(M_{\rm peak}\approx 3.9\) for the single-body model to \(M_{\rm peak}\approx 3.1\) for the fragmentation cases, corresponding to an increase of $\sim$ 0.8~mag. These results suggest that fragmentation concentrates luminosity into a shorter trail with corresponding brighter peak magnitude, whereas the single-body approximation distributes the emission smoothly along a much longer predicted trajectory resulting in meteors systematically more extended and fainter.

The meteoroid sizes represented in the observational sample, 0.4--10~mm, overlap the range in which substantial ablation is expected at Mars \citep{Pabari2023_venusMars} and which contributes efficiently to optical meteor production. The resulting Martian light-curve predictions therefore provide an observationally informed, fragmentation-based estimate of the peak magnitudes and luminous extents of meteoroids within this size range.

Although optical meteors have not yet been confirmed at Mars, the predicted ablation region for mm-sized sporadic meteoroids overlaps the $\sim$65--100~km altitude range in which Mg and Fe ion layers have been reported in the Martian ionosphere \citep{Molina2008MetallayersMARSVENUSTITAN}. This agreement provides an independent consistency check on the fragmentation-inclusive framework. The resulting altitude--brightness distributions offer observation-ready guidance for future surface-based instruments designed to detect optical meteors at Mars. Such detections would constrain variations in the meteoroid environment across heliocentric distance and planetary atmospheres and improve population-based environment models, including NASA MEM and ESA IMEM, by linking predicted fluxes and size--speed distributions to observable atmospheric-entry signatures.

\section*{Acknowledgments}

Funding for this work was provided by the NASA Meteoroid Environment Office under cooperative agreement 80NSSC24M0060 and the European Space Agency (ESA) through Contract Number 4000145350, as part of the ESA Initial Support for Innovation (EISI) program.
The authors would like to thank Dr. Bill Cooke, Dr. Mark Millinger and Dr. Daeyoung Lee for providing insight and expertise that assisted the research.

\section*{Note on Code Availability}

Implementation of all methods used in this work is published as open source on the following GitHub web page:
\begin{itemize}
    \item WesternMeteorPyLib: \\\url{https://github.com/wmpg/WesternMeteorPyLib}
\end{itemize}
Readers are encouraged to contact the lead author in the event they are not able to obtain the code on-line.

\section*{Declaration of generative AI and AI-assisted technologies in the writing process}

During the preparation of this work the authors used ChatGPT in order to improve language and readability. After using this tool/service, the authors reviewed and edited the content as needed and take full responsibility for the content of the publication.



\appendix

\section{Dynamic nested sampling model}\label{sec:AppendixDynami_nested}

To infer meteoroid physical properties from the Earth observations, we applied dynamic nested sampling (DNS) to each of the 144 events, following the approach described by \citet{vovk2026inferring}. DNS provides posterior samples and an estimate of the Bayesian evidence, which quantifies how well a given model explains the observations. By adaptively concentrating computational effort in the highest-likelihood and most informative regions of parameter space, DNS remains efficient for multi-modal and highly covariant posterior distributions while providing robust uncertainty estimates. We used the \texttt{dynesty} implementation of \cite{speagle2020dynesty} to explore the model parameter space and to quantify uncertainties and parameter correlations.

The forward model was based on the erosion--fragmentation formulation of \citet{borovivcka2007atmospheric}, coupled with the luminous efficiency model of \citet{vida2024first}. The luminous efficiency was evaluated as a function of both velocity and mass for the main body and for released grains. We adopted a fixed zenith angle from the geometric trajectory solution, a grain density of $3000$~kg\,m$^{-3}$, a drag coefficient $\Gamma = 1$, and a shape factor $A = 1.21$. All entry simulations were initiated at an altitude of 180~km to ensure a consistent definition of the initial state vector across events.

Assuming independent Gaussian errors, the log-likelihood for $N$ data points is
\begin{equation}
\label{eq:log_likelihood}
\log\mathcal{L}\left(\boldsymbol{\Theta}\right)
=
-\frac{1}{2}\sum_{i=1}^{N}
\left[
\frac{\left(x_i-\mu_i\left(\boldsymbol{\Theta}\right)\right)^2}{\sigma^2}
+\log\!\left(2\pi\sigma^2\right)
\right],
\end{equation}
where $x_i$ are the observations, $\mu_i(\boldsymbol{\Theta})$ the corresponding model predictions, and $\sigma$ the assumed uncertainties.

For our meteor dataset, the total likelihood was defined as the sum of separate contributions from the photometric and dynamical constraints,
\begin{equation}
\log \mathcal{L}(\boldsymbol{\Theta})
=
\log \mathcal{L}_{\mathrm{lum}}(\boldsymbol{\Theta})
+
\log \mathcal{L}_{\mathrm{lag}}(\boldsymbol{\Theta}),
\end{equation}
where both $\log \mathcal{L}_{\mathrm{lum}}$ and $\log \mathcal{L}_{\mathrm{lag}}$ follow Eq.~\ref{eq:log_likelihood}. We treated the effective scatter in each dataset as a free parameter by setting $\sigma_{\mathrm{lum},i}=\sigma_{\mathrm{lum}}$ and $\sigma_{\mathrm{lag},i}=\sigma_{\mathrm{lag}}$ for all $i$. The parameters $\sigma_{\mathrm{lum}}$ and $\sigma_{\mathrm{lag}}$ therefore act as uncertainty-inflation terms that absorb unmodelled variance and enable a balanced fit to both the light curve and the lag profile. Here, the lag quantifies the meteoroid deceleration by measuring how far the observed meteor falls behind a reference trajectory with constant initial velocity. 

All 144 DNS inversions were performed on a dual-socket Linux server equipped with two Intel Xeon Silver 4214 processors operating at 2.20~GHz. Each processor contains 12 physical cores with two hardware threads per core, providing a total of 24 physical CPU cores and 48 logical processing threads. The system has approximately 192~GB of RAM (187~GiB usable memory). The runtime of each inversion depended primarily on the number of free parameters in the model, with more complex parameterizations requiring longer convergence times, but typically was of order several hundred total CPU-hours.



\subsection{Erosion--fragmentation model parameterisation}

The erosion--fragmentation model included two erosion stages, allowing the meteoroid to change its effective material properties as fragmentation progressed. The set of free parameters, $\boldsymbol{\Theta}$, was
\[
\boldsymbol{\Theta} =
\begin{bmatrix}
v_0,\; m_{0},\; \rho,\; \sigma,\; h_{e},\; \eta,\\
s,\; m_{l},\; m_{u},\; h_{e2},\; \rho_2,\; \sigma_2,\\
\eta_2,\; \sigma_{\mathrm{lag}},\; \sigma_{\mathrm{lum}}
\end{bmatrix},
\]
where $v_0$ is the initial velocity at 180~km and $m_0$ is the initial meteoroid mass. The bulk density in the first erosion stage is $\rho$. The ablation coefficient $\sigma$ governs mass loss by vaporisation, while the erosion coefficient $\eta$ governs mass transfer from the parent body into released grains. Grain release was described by a power-law mass distribution with mass index $s$ and minimum and maximum grain masses $m_l$ and $m_u$. The onset of grain release occurred at the erosion height $h_e$.

A second erosion stage was activated at height $h_{e2}$, after which the meteoroid material properties were allowed to change to $(\rho_2,\sigma_2,\eta_2)$. We enforced $\rho_2 \ge \rho$ to reflect the possibility that a more compact component (e.g., a core) becomes dominant after the initial erosion phase. Finally, $\sigma_{\mathrm{lag}}$ and $\sigma_{\mathrm{lum}}$ were treated as nuisance parameters describing the effective scatter (uncertainty inflation) in the lag and luminosity likelihood terms, respectively. Each dynesty run using 15 variables on our cluster takes on average 18 hours of node time. The full set of prior distributions is listed in Table~\ref{tab:priors_1frg}.

\begin{table}
\caption{Adopted prior distributions used as inputs for the erosion--fragmentation model, following \citet{vovk2026inferring}. Here, $h_{\text{beg}}$ denotes the beginning height, $h_{\text{end}}$ the end height, $h_{\text{peak}}$ the height of the light-curve peak, and $\mathcal{O}$ the order of magnitude derived from the photometric mass estimate $m_{\text{phot.}}$ for a luminous efficiency of $0.7\%$.}
\label{tab:priors_1frg}
\begin{tabular*}{\tblwidth}{@{}llp{4cm}@{}}
\toprule
\textbf{Parameter} & \textbf{Prior type} & \textbf{Range or formula} \\ 
\midrule
$v_0$ [km/s] & Gaussian & Mean at 100 m/s over first-frame speed, $\sigma = 500$ m/s \\
$m_{0}$ [kg] & Uniform & $0.1\,\mathcal{O}(m_{\text{phot.}})$ -- $20\,\mathcal{O}(m_{\text{phot.}})$ \\
$\rho$ [kg/m$^{3}$] & Uniform & 100 -- 4000 \\
$\sigma$ [kg/MJ] & Uniform & 0.001 -- 0.05 \\
$h_{e}$ [km] & Uniform & $(\frac{1}{2}h_{\text{beg}} + \frac{1}{2}h_{\text{peak}})$ -- $(\frac{3}{2}h_{\text{beg}} - \frac{1}{2}h_{\text{peak}})$ \\
$\eta$ [kg/MJ] & log$_{10}$-uniform & 0 -- 1 \\
$s$ & Uniform & 1 -- 3 \\
$m_{l}$ [kg] & log$_{10}$-uniform & $5\times10^{-12}$ -- $1\times10^{-9}$ \\
$m_{u}$ [kg] & log$_{10}$-uniform & $\max\left(m_{l}, 1\times10^{-10}\right)$ -- $1\times10^{-7}$ \\
$h_{e2}$ [km] & Uniform & $h_{\text{end}}$ -- $\min\left(h_e,\, h_{\text{beg}}\right)$ \\
$\rho_{2}$ [kg/m$^{3}$] & Uniform & $\rho$ -- 4000 \\
$\sigma_{2}$ [kg/MJ] & Uniform & 0.001 -- 0.05 \\
$\eta_{2}$ [kg/MJ] & log$_{10}$-uniform & 0 -- 1 \\
$\sigma_{\mathrm{lag}}$ [m] & Inverse Gamma & Mode from lag fit, $\alpha = 10$ \\
$\sigma_{\mathrm{lum}}$ [W] & Inverse Gamma & Mode from lum fit, $\alpha = 5$ \\
\bottomrule
\end{tabular*}
\end{table}

\subsection{Single-body ablation model parameterisation}

To provide a baseline consistent with the standard approach used in many planetary meteor studies, we also fitted each event using a classical single-body ablation model \citep{opik1958physics}. In this formulation, the meteoroid is treated as an intact body that undergoes smooth mass loss by ablation, without explicit grain release or fragmentation. We retained the same fixed parameters as in the erosion--fragmentation model (zenith angle, grain density where relevant, $\Gamma$, and $A$), and initiated simulations at 180~km.

The free parameter vector for the single-body model was
\[
\boldsymbol{\Theta} =
\begin{bmatrix}
v_0,\; m_{0},\; \rho,\; \sigma,\; \sigma_{\mathrm{lag}},\; \sigma_{\mathrm{lum}}
\end{bmatrix},
\]
where $v_0$ and $m_0$ are the initial velocity and mass at 180~km, $\rho$ is the bulk density of the intact meteoroid, and $\sigma$ is the ablation coefficient controlling the rate of mass loss by vaporization. As above, $\sigma_{\mathrm{lag}}$ and $\sigma_{\mathrm{lum}}$ represent nuisance terms capturing the effective scatter in the lag and luminosity likelihood components. Compared to the fit given by the erosion fragmentation model the single body ablation fit are worse in both light-curve and dynamic fit. Each dynesty run using 6 variables on our cluster takes on average 30 minutes. The adopted prior distributions are given in Table~\ref{tab:priors_1body}.

\begin{table}
\caption{Adopted prior distributions used as inputs for the single-body ablation model.}
\label{tab:priors_1body}
\begin{tabular*}{\tblwidth}{@{}llp{4cm}@{}}
\toprule
\textbf{Parameter} & \textbf{Prior type} & \textbf{Range or formula} \\
\midrule
$v_0$ [km/s] & Gaussian & Mean at 100 m/s over first-frame speed, $\sigma = 500$ m/s \\
$m_{0}$ [kg] & Uniform & $0.1\,\mathcal{O}(m_{\text{phot.}})$ -- $20\,\mathcal{O}(m_{\text{phot.}})$ \\
$\rho$ [kg/m$^{3}$] & Uniform & 100 -- 4000 \\
$\sigma$ [kg/MJ] & Uniform & 0.001 -- 0.05 \\
$\sigma_{\mathrm{lag}}$ [m] & Inverse Gamma & Mode from lag fit, $\alpha = 10$ \\
$\sigma_{\mathrm{lum}}$ [W] & Inverse Gamma & Mode from lum fit, $\alpha = 5$ \\
\bottomrule
\end{tabular*}
\end{table}


For mm-sized meteoroids, the single-body model is not expected to reproduce the full observed light-curve and deceleration behaviour because fragmentation is common in this size regime \citep{Subasinghe2016}. Consistent with this expectation, the single-body fits in our dataset generally provided poorer joint agreement with the photometric and dynamical constraints than the erosion--fragmentation model, and the inferred parameters are therefore more susceptible to bias when fragmentation is neglected, for example through inflated ablation coefficients and underestimated bulk densities \citep{popova2019modelling}. 


\section{Mars atmospheric model}\label{sec:MarsAtmosphere}

Accurate atmospheric density profiles were required because the ablation, deceleration, and luminosity depend sensitively on $\rho(h)$. For Earth, we used the NRLMSIS-00 model to evaluate the atmospheric density at the measured meteor locations and times. Following \citet{Vida2021}, the density profile used within the iterative modelling was represented by a 7th-order polynomial fit to reduce computational cost while preserving the local atmospheric structure relevant for the trajectory.

For Mars, the meteor geographic location and local time were not known a priori. We therefore adopted a representative background atmosphere from the Mars Climate Database v6.1 (MCD\_v6.1) \citep{forget1999improvedmars, millour2009mars}, using a climatology average solar scenario with diurnal-mean conditions. Specifically, we used $L_s = 160.3^{\circ}$, latitude $0.0^{\circ}$~N, longitude $0.0^{\circ}$~E, and altitude defined relative to the MCD reference surface (ALS), with a diurnal mean over all local times. As done for Earth, we fit a 7th-order polynomial to the resulting density profile to accelerate the forward-model evaluations.

\begin{figure}
\centering
\includegraphics[width=\columnwidth]{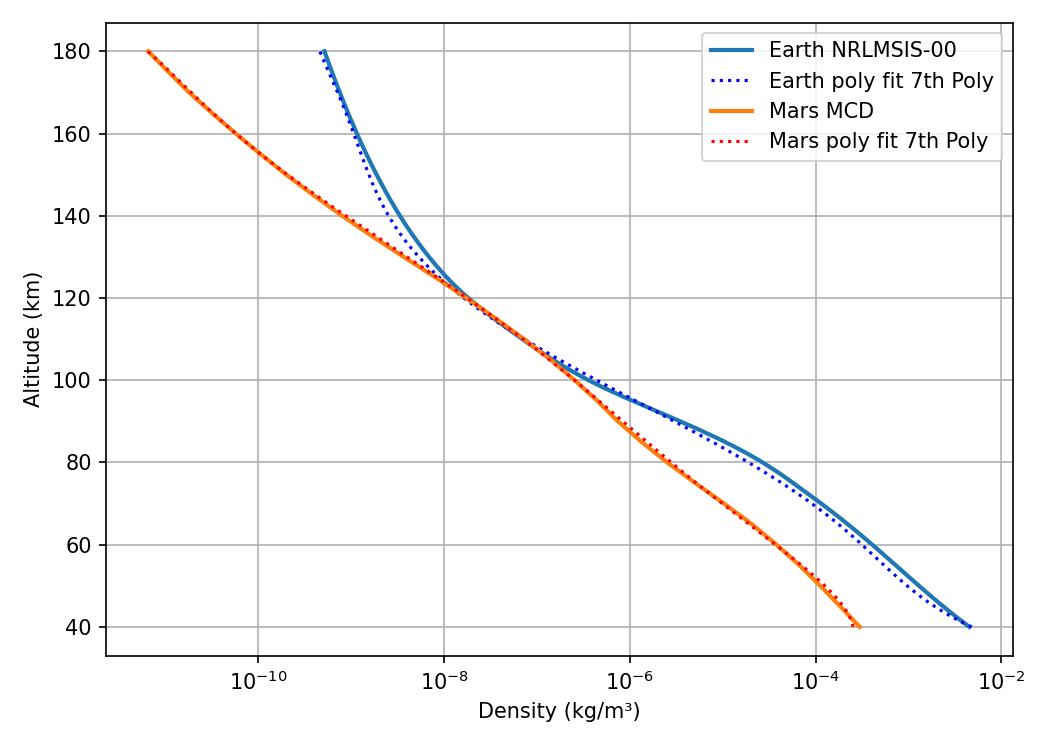}
\caption{Atmospheric density profiles used in the meteor simulations on Earth (NRLMSIS-00) and Mars (MCD\_v6.1; diurnal-mean climatology case). Dotted curves show the corresponding 7th-order polynomial fits adopted to accelerate the density evaluations during the iterative entry modelling.}
\label{img:atmosp}
\end{figure}

The Earth atmosphere is denser than the Martian atmosphere over most altitudes relevant for meteor ablation. However, the two profiles exhibited an overlap region (approximately 100--120~km above the reference level), which affected the relative ablation and luminosity heights predicted on the two planets.

\section{Projecting Intercept Velocities}\label{sec:speed_mars}

To project the potential impact parameters of an observed Earth-crossing meteoroid onto a hypothetical intercept with Mars, we employ a 3D orbital mechanics model. This model uses the five standard heliocentric Keplerian orbital elements of the meteoroid, as determined from its Earth-based observations: semi-major axis $a$, eccentricity $e$, inclination $i$, argument of perihelion $\omega$, and longitude of the ascending node $\Omega$. The procedure first validates the Earth intercept by finding the closest orbital node and then calculates the two possible intercept scenarios at Mars's mean orbital radius.

The core of the calculation is the conversion of the orbital elements into 3D heliocentric ecliptic state vectors (position $\vec{r}$ and velocity $\vec{v}$). This requires a sixth parameter, the true anomaly ($\nu$), which defines the object's position along its orbit.

First, the semi-latus rectum $p$ is calculated:
\begin{equation}
    p = a(1 - e^2)
    \label{eq:semi_latus_rectum}
\end{equation}
The heliocentric distance $r$ for a given true anomaly $\nu$ is then:
\begin{equation}
    r = \frac{p}{1 + e \cos(\nu)}
    \label{eq:radius}
\end{equation}

The object's state vectors in the perifocal (orbital) frame ($\hat{P}, \hat{Q}, \hat{W}$) are given by:
\begin{equation}
    \vec{r}_{PQW} = \begin{bmatrix}
        r \cos(\nu) \\
        r \sin(\nu) \\
        0
    \end{bmatrix}
    \label{eq:r_perifocal}
\end{equation}
\begin{equation}
    \vec{v}_{PQW} = \sqrt{\frac{\mu_{Sun}}{p}} \begin{bmatrix}
        -\sin(\nu) \\
        e + \cos(\nu) \\
        0
    \end{bmatrix}
    \label{eq:v_perifocal}
\end{equation}
where $\mu_{Sun}$ is the standard gravitational parameter of the Sun.

These vectors are rotated into the heliocentric ecliptic frame ($XYZ$) using a rotation matrix $R$:
\begin{equation}
    R =
    \begin{bmatrix}
        c_{\Omega}c_{\omega} - s_{\Omega}s_{\omega}c_{i} & -c_{\Omega}s_{\omega} - s_{\Omega}c_{\omega}c_{i} &  s_{\Omega}s_{i} \\
        s_{\Omega}c_{\omega} + c_{\Omega}s_{\omega}c_{i} & -s_{\Omega}s_{\omega} + c_{\Omega}c_{\omega}c_{i} & -c_{\Omega}s_{i} \\
        s_{\omega}s_{i}                                &  c_{\omega}s_{i}                                &  c_{i}
    \end{bmatrix}
    \label{eq:rotation_matrix}
\end{equation}
where $c_{\alpha} = \cos(\alpha)$ and $s_{\alpha} = \sin(\alpha)$. The final state vectors are:
\begin{equation}
    \vec{r}_{ecl} = R \cdot \vec{r}_{PQW} \quad \text{and} \quad \vec{v}_{ecl} = R \cdot \vec{v}_{PQW}
    \label{eq:state_vectors}
\end{equation}

\begin{figure}
\centering
\includegraphics[width=\columnwidth]{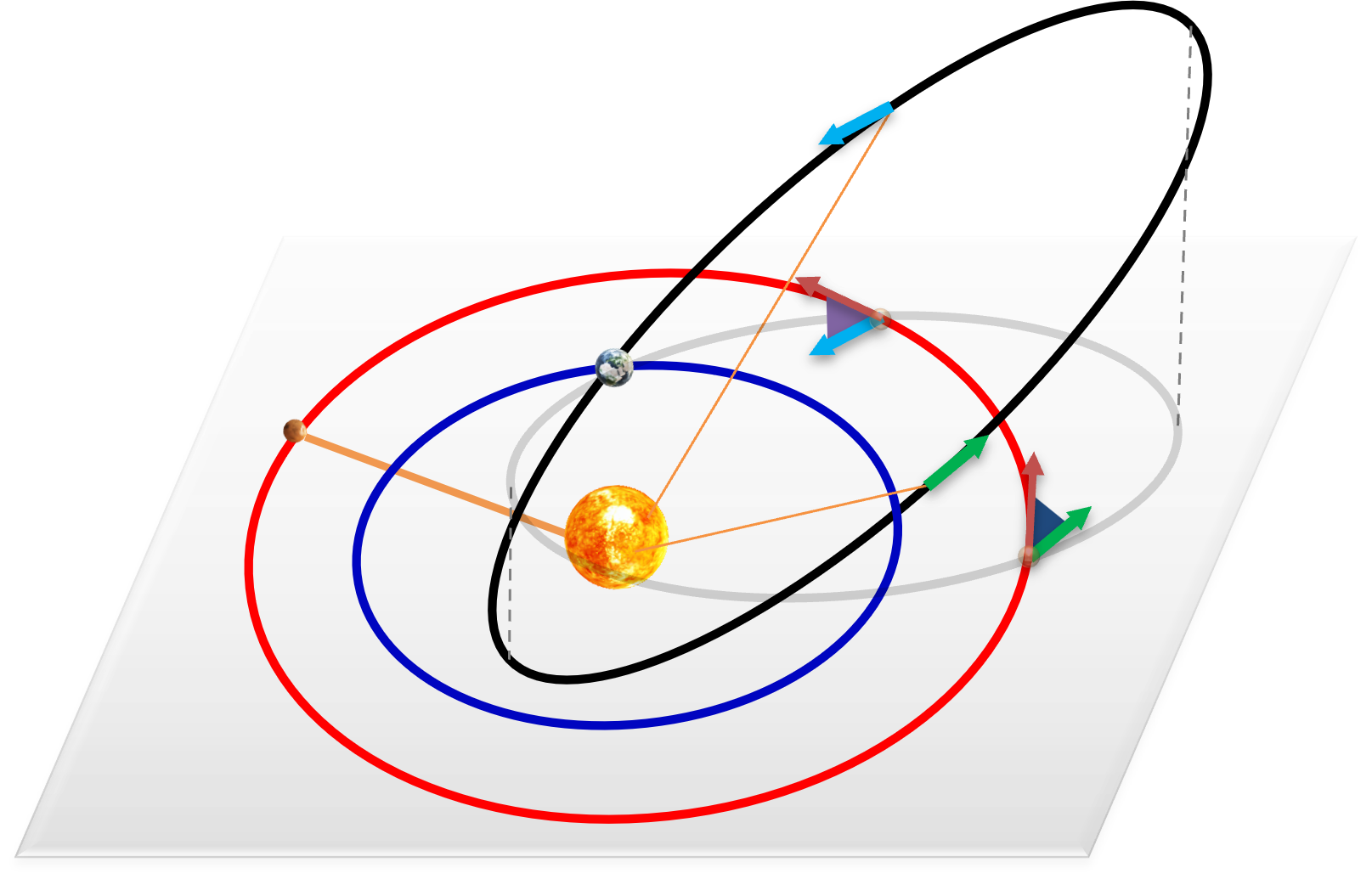}
\caption{Schematic of the 3D orbital-geometry model used to project an Earth-crossing meteoroid onto a hypothetical Mars intercept. The blue curve shows the (stylised) orbit of Earth, the red curve shows the orbit of Mars, and the black curve is the meteoroid’s true heliocentric orbit. The grey curve is the same orbit projected into the ecliptic plane ($i = 0^\circ$). Velocity vectors are drawn at the two radii where the projected orbit crosses Mars’s orbital distance, illustrating how different encounter angles lead to different intercept speeds.}
\label{img:mars_projected_speed} 
\end{figure}

For Mars, we project a hypothetical intercept at its mean orbital radius ($r_{Mars} \approx 1.524$ AU). We first determine the true anomalies where the meteoroid's orbit crosses this radius by solving Eq. \ref{eq:radius} for $\nu$:
\begin{equation}
    \theta_{Mars} = \pm \arccos \left( \frac{p - r_{Mars}}{e \cdot r_{Mars}} \right)
    \label{eq:nu_mars}
\end{equation}
This yields two possible scenarios, provided $r_{Mars}$ is between the orbit's perihelion and aphelion:
\begin{enumerate}
    \item {Inbound Intercept} ($\theta_{Mars} < 0$): The object is traveling inwards toward the Sun.
    \item {Outbound Intercept} ($\theta_{Mars} > 0$): The object has passed perihelion and is traveling outwards.
\end{enumerate}
For each scenario, the corresponding true anomaly is used to find the object's state vectors ($\vec{r}_{obj, M}$, $\vec{v}_{obj, M}$) at the Mars-crossing point, a simple diagram is shown in Figure \ref{img:mars_projected_speed}.

For both Earth and Mars, the atmospheric entry speed ($v_i$) is found by considering the object's hyperbolic (relative to the planet) encounter with the planet. We first compute the planet's velocity vector ($\vec{v}_{planet}$) at the intercept position, assuming a circular, prograde orbit in the ecliptic plane. The relative velocity vector at encounter, or hyperbolic excess velocity ($\vec{v}_{\infty}$), is:
\begin{equation}
    \vec{v}_{\infty} = \vec{v}_{obj} - \vec{v}_{planet}
    \label{eq:v_infinity}
\end{equation}

\begin{figure}
\centering
\includegraphics[width=\columnwidth]{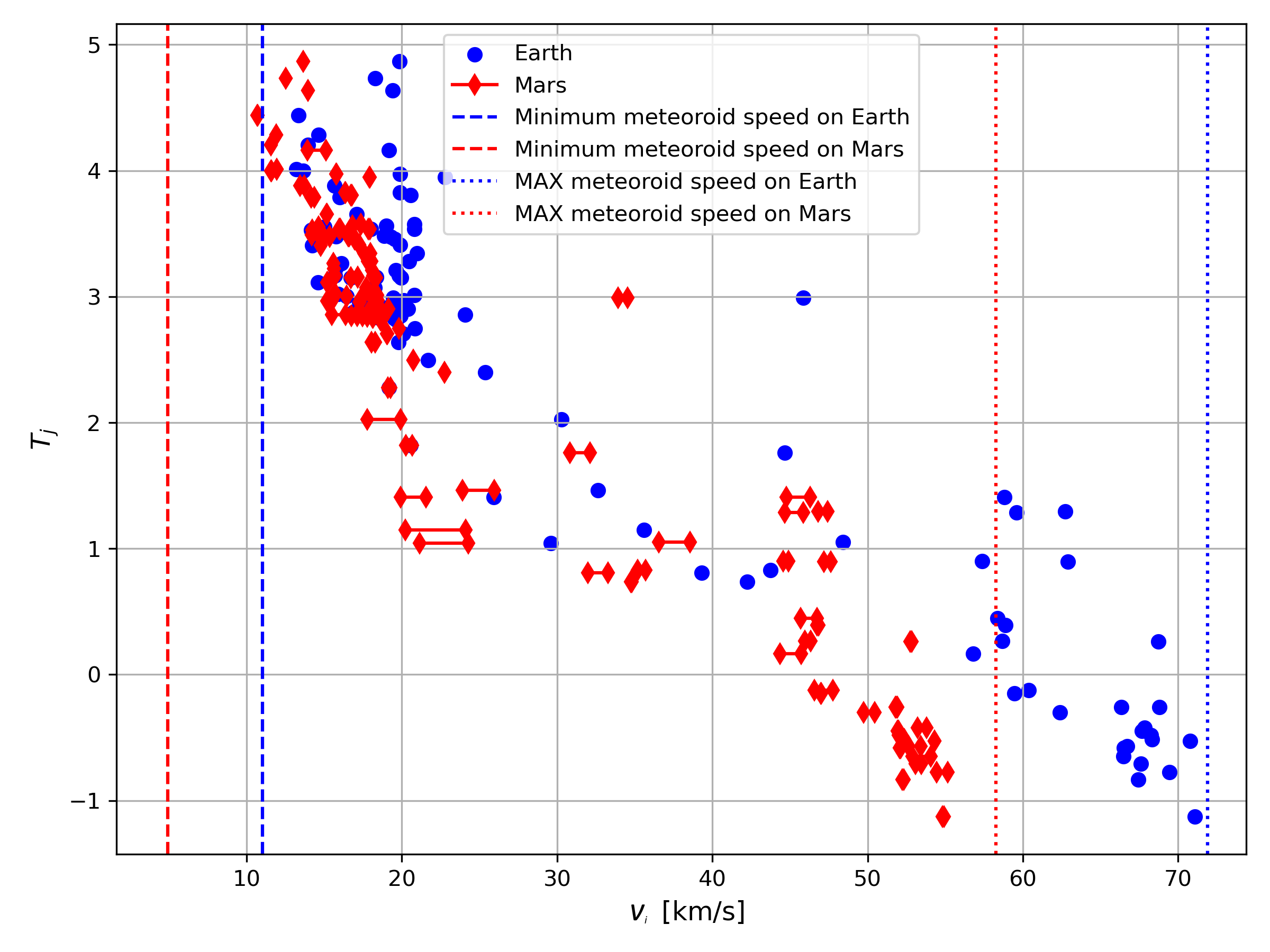}
\caption{Atmospheric-entry speed at Earth (blue) and the corresponding range of possible entry speeds at Mars (red) for Tisserand's parameter. The Mars interval spans the relative speeds obtained for the two admissible Mars intercept geometries (inbound and outbound solutions), which differ in the angle between the meteoroid’s velocity and Mars’s orbital motion.}
\label{img:mars_Vinf} 
\end{figure}

From the conservation of energy, the squared entry speed $v_i^2$ at a given atmospheric altitude $h$ is the sum of the squared hyperbolic excess speed and the squared escape velocity ($v_{esc}$) from that altitude:
\begin{equation}
    v_i^2 = v_{\infty}^2 + v_{esc}^2
    \label{eq:entry_speed}
\end{equation}
where $v_{\infty} = \|\vec{v}_{\infty}\|$ and $v_{esc}$ is defined as:
\begin{equation}
    v_{esc} = \sqrt{\frac{2 G M_{planet}}{R_{planet} + h}}
    \label{eq:escape_velocity}
\end{equation}
Here, $G$ is the gravitational constant, $M_{planet}$ is the mass of the planet, $R_{planet}$ is its mean radius, and $h$ is the atmospheric entry altitude. To make the initial conditions comparable between planets, we chose the Mars entry altitude $h=140$ km such that the local atmospheric density matches the density at $h=180$ km on Earth (the start altitude of the Earth simulations). This $v_i$ represents the final computed speed of the object as it enters the planet's atmosphere. The different speed on Earth and the resulting speed at Mars is shown in Figure \ref{img:mars_Vinf}.

Of the two estimated entry speeds, $v_i$, obtained for each simulation, we adopted their mean as the representative entry speed at Mars. This value was used as the initial velocity for the Martian meteoroid simulations.

\begin{figure*}
\centering
\includegraphics[width=0.9\textwidth]{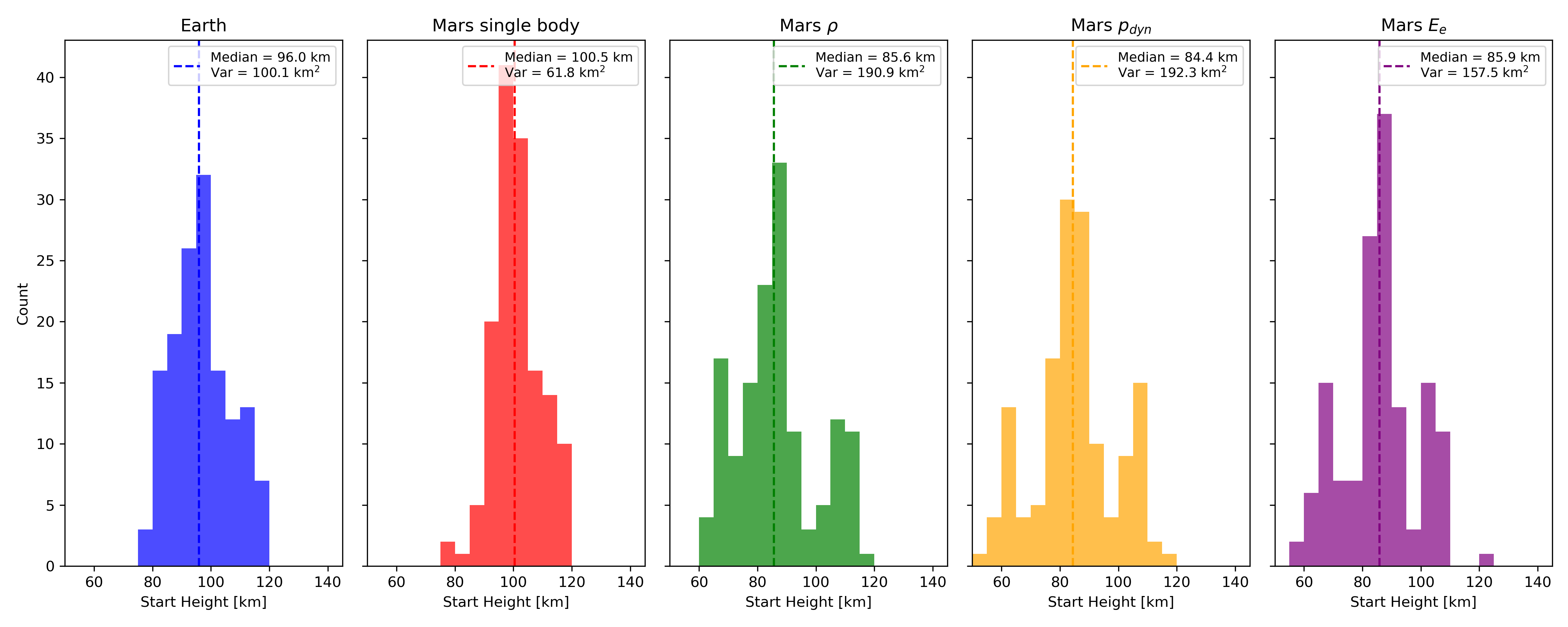}
\includegraphics[width=0.9\textwidth]{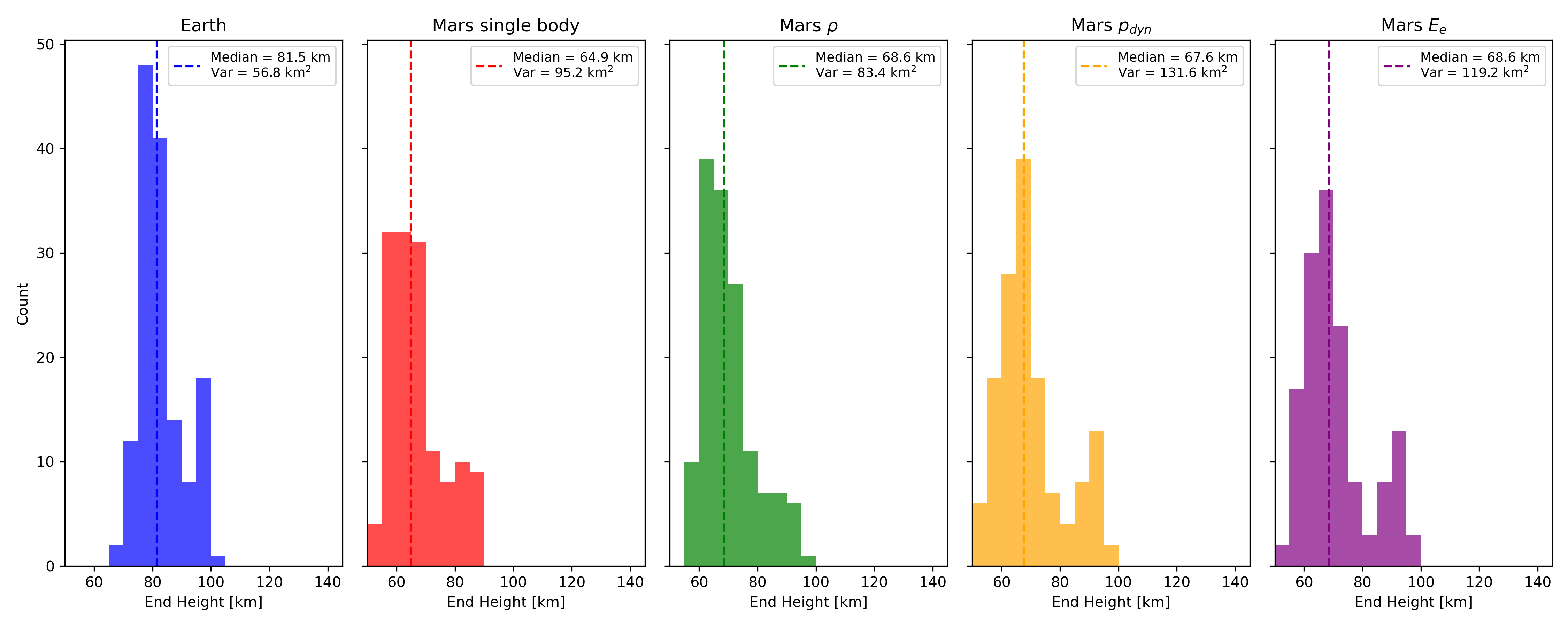}
\includegraphics[width=0.9\textwidth]{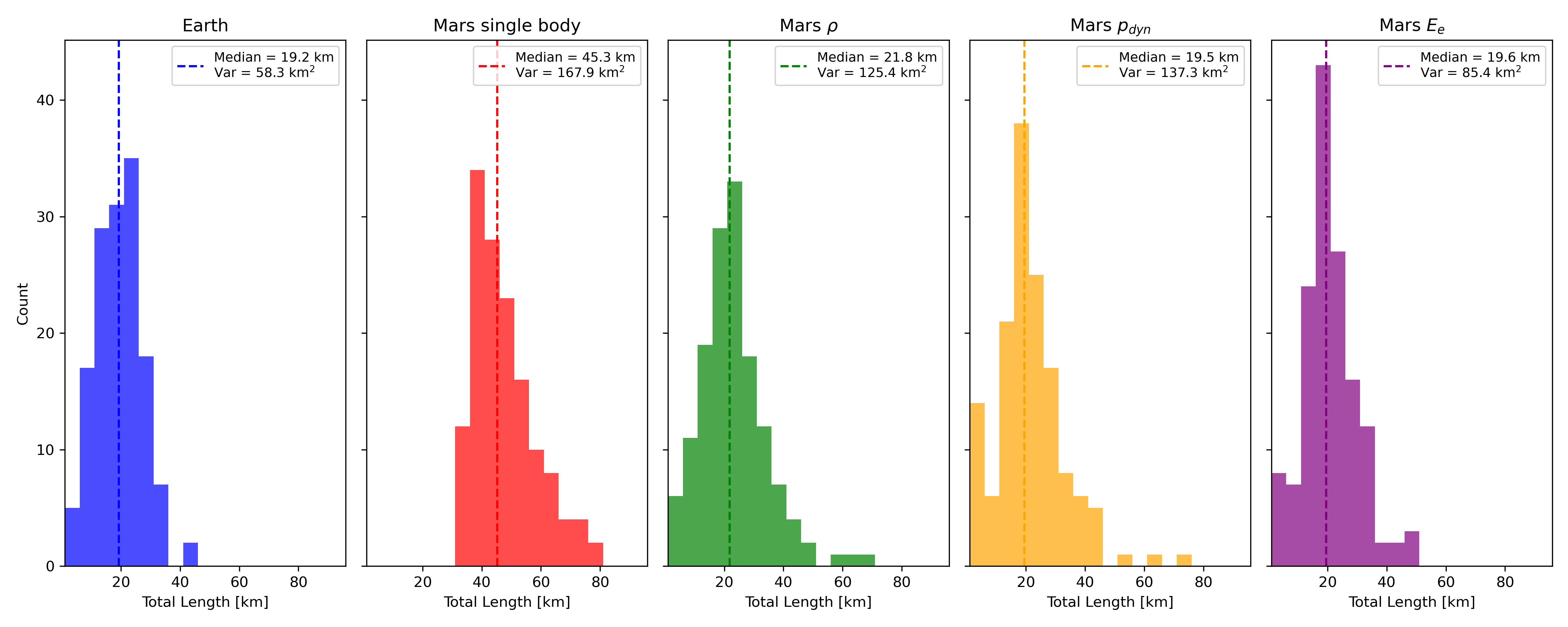}
\caption{Distributions of the meteor luminous properties obtained using the common \(M_{\rm peak}+2.5\) detection threshold. From top to bottom, the panels show the luminous beginning height, luminous ending height, and along-track trail length for the Earth erosion--fragmentation simulations, the single-body cases and the Martian atmospheric-density, dynamic-pressure, accumulated-energy cases. Median values and variances are indicated for each distribution. The Martian single-body model predicts systematically higher beginning heights, lower ending heights, and substantially longer luminous trails than the erosion--fragmentation cases.}\label{fig:all_height_length}
\end{figure*}

\section{Mapping erosion-onset conditions from Earth to Mars}
\label{sec:erosion_trigger_mapping}

The erosion heights inferred from the Earth observations cannot be transferred directly to Mars because the atmospheric density profiles and entry velocities differ between the two planets. Instead, the physical condition reached at each terrestrial erosion height, \(h_e\) and \(h_{e2}\), was calculated and matched to the corresponding condition along the simulated Martian trajectory.

For the atmospheric-density trigger, the Martian erosion height was defined by
\begin{equation}
\rho_{\mathrm{atm,M}}\left(h_{e,\mathrm{M}}\right)
=
\rho_{\mathrm{atm,E}}\left(h_{e,\mathrm{E}}\right),
\end{equation}
where the subscripts E and M denote Earth and Mars, respectively. Similarly, for the dynamic-pressure trigger,
\begin{equation}
\rho_{\mathrm{atm,M}}\left(h_{e,\mathrm{M}}\right)
v_{\mathrm{M}}^2\left(h_{e,\mathrm{M}}\right)
=
\rho_{\mathrm{atm,E}}\left(h_{e,\mathrm{E}}\right)
v_{\mathrm{E}}^2\left(h_{e,\mathrm{E}}\right).
\end{equation}
The same procedure was applied independently to the first and second erosion stages.

The accumulated-energy trigger required integration of the aerodynamic energy received before erosion. Assuming no substantial change in mass or velocity over an integration interval, the received energy per unit cross-sectional area was calculated as
\begin{equation}
\mathcal{E}_{A}(h)
=
\frac{\Lambda v^2}{2\cos z}
\int_{h}^{h_0}
\rho_{\mathrm{atm}}(h')\,dh',
\end{equation}
where \(\Lambda\) is the heat-transfer coefficient, assumed here to be unity; (z) is the meteoroid zenith angle; and \(h_0\) is the starting altitude of the simulation. For the Martian simulations, \(h_0\) was selected as the altitude at which the atmospheric density equals that at 180~km on Earth, corresponding to the initial altitude adopted for the terrestrial simulations. The corresponding energy received per unit meteoroid mass was
\begin{equation}
\mathcal{E}_{m}(h)
=
\mathcal{E}_{A}(h)
A\rho_{\mathrm{m}}^{-2/3}m^{-1/3},
\end{equation}
where \(A\) is the shape factor, \(\rho_{\mathrm{m}}\) is the meteoroid bulk density, and \(m\) is its instantaneous mass.

In the numerical implementation, the trajectory was divided into intervals over which the mass and velocity were approximately constant. The energy received in each interval was calculated using its instantaneous mass and velocity, and the contributions were summed to obtain the cumulative received energy,
\begin{equation}
E_{\mathrm{e}}(h_n)
=
\sum_{j=1}^{n}
\mathcal{E}_{m,j}\,m_j .
\end{equation}
The Martian erosion height was then selected as the altitude at which the cumulative energy most closely matched the value inferred at the corresponding terrestrial erosion height:
\begin{equation}
E_{\mathrm{e,M}}\left(h_{e,\mathrm{M}}\right)
=
E_{\mathrm{e,E}}\left(h_{e,\mathrm{E}}\right).
\end{equation}
When two erosion stages were present, this matching was performed separately for \(h_e\) and \(h_{e2}\).

The accumulated energy represents an integrated aerodynamic-energy input and is used only as a phenomenological erosion-onset condition; it does not explicitly model internal heat conduction or the physical properties of any binding material. 

\section{Detection threshold and luminous-trail calculation}
\label{sec:detection_threshold}

To compare the simulated meteors consistently, the detectable portion of every light curve was defined relative to its peak brightness. For each Earth and Mars simulation, the peak absolute magnitude was
\begin{equation}
M_{\rm peak}=\min\left[M(t)\right].
\end{equation}
The meteor was considered detectable at all time steps satisfying
\begin{equation}
M(t)\leq M_{\rm peak}+2.5.
\end{equation}
Thus, the adopted threshold includes the portion of the light curve extending from the peak to \(2.5\)~mag fainter than the peak. This relative threshold was applied identically to the Earth erosion--fragmentation simulations, the three Martian erosion--fragmentation cases, and the Martian single-body case. The Earth beginning and ending heights were therefore recomputed from the simulated light curves using the same criterion adopted for Mars, rather than using the observational camera limits. This allows the luminous morphology of all cases to be compared over the same relative magnitude range.

The position of the meteor along its trajectory was defined using the leading active fragment. For each fragment \(j\), the accumulated along-track distance was updated at every integration step according to
\begin{equation}
\ell_j(t+\Delta t)
=
\ell_j(t)+v_j(t)\Delta t,
\end{equation}
where \(v_j\) is the instantaneous fragment velocity. At each time step, the leading fragment was selected as the active fragment with the greatest accumulated trajectory length,
\begin{equation}
j_{\rm lead}(t)
=
\underset{j\in\mathcal{A}(t)}{\operatorname{arg\,max}}
\left[\ell_j(t)\right],
\end{equation}
where \(\mathcal{A}(t)\) is the set of active fragments. Its length and height were stored in the arrays
\(\ell_{\rm lead}(t)\) and \(h_{\rm lead}(t)\), corresponding to the leading fragment length and
the leading fragment height, respectively. In the erosion--fragmentation simulations, this procedure follows the frontmost surviving fragment or grain, whereas in the single-body simulations it follows the intact main body.

Let \(t_{\rm beg}\) and \(t_{\rm end}\) be the first and last time steps satisfying the \(2.5\)-mag criterion. The luminous beginning and ending heights were calculated as
\begin{equation}
h_{\rm beg}=h_{\rm lead}(t_{\rm beg}),
\qquad
h_{\rm end}=h_{\rm lead}(t_{\rm end}),
\end{equation}
and the along-track luminous trail length was
\begin{equation}
L
=
\ell_{\rm lead}(t_{\rm end})
-
\ell_{\rm lead}(t_{\rm beg}).
\end{equation}

The magnitude threshold was applied to the total simulated luminosity, including contributions from the main body and all released grains, while the heights and along-track distances were assigned using the leading-fragment trajectory.

The resulting distributions are shown in Fig.~\ref{fig:all_height_length}. The median Martian beginning heights are \(84.4\)--\(85.9\)~km for the three erosion--fragmentation cases, compared with \(100.5\)~km for the single-body model and \(96.0\)~km for the Earth simulations. The single-body model therefore begins approximately \(15\)~km higher than the Martian fragmentation cases. Its median ending height is also lower, at \(64.9\)~km compared with \(67.6\)--\(68.6\)~km for the fragmentation cases and \(81.5\)~km on Earth. Consequently, the median vertical luminous extent is approximately \(35.6\)~km for the single-body model, compared with only \(16.8\)--\(17.3\)~km when fragmentation is included and \(14.5\)~km on Earth.

The strongest difference is found in the along-track trail lengths shown in the bottom panel of Fig.~\ref{fig:all_height_length}. The single-body model produces a median trail length of \(45.3\)~km, more than twice the \(19.5\)--\(21.8\)~km obtained for the three Martian erosion--fragmentation cases and the \(19.2\)~km median obtained for Earth. The single-body trail-length distribution also has the largest variance, \(167.9~\mathrm{km^2}\), compared with \(85.4\)--\(137.3~\mathrm{km^2}\) for the Martian fragmentation cases and \(58.3~\mathrm{km^2}\) for Earth. The single-body approximation on Mars therefore produces luminous trails that are both systematically longer and more broadly distributed than the Martian meteor simulations using the erosion fragmentation model. 

\printcredits

\bibliographystyle{cas-model2-names}

\bibliography{cas-refs}

@book{lissauer2019fundamentalplanetary,
  title={Fundamental Planetary Science: Physics, Chemistry and Habitability},
  author={Lissauer, J.J. and de Pater, I.},
  isbn={9781108411981},
  year={2019},
  publisher={Cambridge University Press}
}

@article{forget1999improvedmars,
  title={Improved general circulation models of the Martian atmosphere from the surface to above 80 km},
  author={Forget, Fran{\c{c}}ois and Hourdin, Fr{\'e}d{\'e}ric and Fournier, Richard and Hourdin, Christophe and Talagrand, Olivier and Collins, Matthew and Lewis, Stephen R and Read, Peter L and Huot, Jean-Paul},
  journal={Journal of Geophysical Research: Planets},
  volume={104},
  number={E10},
  pages={24155--24175},
  year={1999},
  publisher={Wiley Online Library}
}

@article{vida2024first,
  title={First holistic modelling of meteoroid ablation and fragmentation: A case study of the Orionids recorded by the Canadian Automated Meteor Observatory},
  author={Vida, Denis and Brown, Peter G and Campbell-Brown, Margaret and Egal, Auriane},
  journal={Icarus},
  volume={408},
  pages={115842},
  year={2024},
  publisher={Elsevier}
}

@article{molina2003meteoric,
  title={Meteoric ions in the atmosphere of Mars},
  author={Molina-Cuberos, Gregorio J and Witasse, Olivier and Lebreton, Jean-Pierre and Rodrigo, Rafael and L{\'o}pez-Moreno, Jos{\'e} J},
  journal={Planetary and Space Science},
  volume={51},
  number={3},
  pages={239--249},
  year={2003},
  publisher={Elsevier}
}

@techreport{millour2009mars,
  title={The mars climate database (version 4.3)},
  author={Millour, Ehouarn and Forget, F and Gonz{\'a}lez-Galindo, F and Spiga, A and Lebonnois, S and Lewis, SR and Montabone, L and Read, PL and Lopez-Valverde, MA and Gilli, G and others},
  year={2009},
  institution={SAE Technical Paper}
}

@article{Pabari2023_venusMars,
   author = {Jayesh P. Pabari and Srirag N. Nambiar and Rashmi and Sonam Jitarwal},
   doi = {10.1016/j.pss.2022.105617},
   issn = {00320633},
   journal = {Planetary and Space Science},
   month = {2},
   publisher = {Elsevier Ltd},
   title = {Metallic ion layers in planetary atmosphere: Boundary conditions and IDP flux},
   volume = {226},
   year = {2023},
}

@article{Crismani2022meteoiricProtonMars,
   author = {Juan Diego Carrillo-Sánchez and Diego Janches and John M.C. Plane and Petr Pokorný and Menelaos Sarantos and Matteo M.J. Crismani and Wuhu Feng and Daniel R. Marsh},
   doi = {10.3847/PSJ/ac8540},
   issn = {26323338},
   issue = {10},
   journal = {Planetary Science Journal},
   pages = {239},
   publisher = {IOP Publishing},
   title = {A Modeling Study of the Seasonal, Latitudinal, and Temporal Distribution of the Meteoroid Mass Input at Mars: Constraining the Deposition of Meteoric Ablated Metals in the Upper Atmosphere},
   volume = {3},
   url = {http://dx.doi.org/10.3847/PSJ/ac8540},
   year = {2022},
}

@article{Vida2021,
   author = {Denis Vida and Peter G. Brown and Margaret Campbell-Brown and Robert J. Weryk and Gunter Stober and John P. McCormack},
   doi = {10.1016/j.icarus.2020.114097},
   issn = {10902643},
   journal = {Icarus},
   month = {1},
   publisher = {Academic Press Inc.},
   title = {High precision meteor observations with the Canadian automated meteor observatory: Data reduction pipeline and application to meteoroid mechanical strength measurements},
   volume = {354},
   year = {2021},
}

@article{Carrillo-Sanchez2019a,
   author = {Juan Diego Carrillo-Sánchez and Juan Carlos Gómez-Martín and David L. Bones and D. Nesvorný and P Pokorný and Mehdi Benna and G.J. Flynn and J.M.C. Plane},
   doi = {10.1016/j.icarus.2019.113395},
   issn = {00191035},
   journal = {Icarus},
   month = {8},
   pages = {113395},
   publisher = {Elsevier Inc},
   title = {Cosmic dust fluxes in the atmospheres of Earth, Mars, and Venus},
   url = {https://doi.org/10.1016/j.icarus.2019.113395 https://linkinghub.elsevier.com/retrieve/pii/S0019103519301824},
   year = {2019},
}

@article{Subasinghe2016,
   author = {D. Subasinghe and M.D. Campbell-Brown and E. Stokan},
   doi = {10.1093/mnras/stw019},
   issn = {13652966},
   issue = {2},
   journal = {Monthly Notices of the Royal Astronomical Society},
   pages = {1289-1298},
   title = {Physical characteristics of faint meteors by light curve and high-resolution observations, and the implications for parent bodies},
   volume = {457},
   year = {2016},
}

@article{borovivcka2007atmospheric,
   author = {J. Borovička and P. Spurný and P. Koten},
   doi = {10.1051/0004-6361},
   issue = {2},
   journal = {Astronomy},
   pages = {661-672},
   publisher = {edpsciences. org},
   title = {Atmospheric deceleration and light curves of Draconid meteors and implications for the structure of cometary dust},
   volume = {473},
   url = {http://www.aanda.org/articles/aa/abs/2007/38/aa8131-07/aa8131-07.html},
   year = {2007},
}

@inproceedings{fleming1993light,
  title={Light curves of faint television meteors},
  author={Fleming, David EB and Hawkes, Robert L and Jones, J},
  booktitle={Meteoroids and their parent bodies},
  pages={261},
  year={1993}
}

@article{vovk2026inferring,
  title={Inferring meteoroid properties with dynamic nested sampling: A case study of orionid and capricornid shower meteors},
  author={Vovk, Maximilian and Brown, Peter G and Vida, Denis and Lee, Daeyoung and Harmos, Emma G},
  journal={Icarus},
  pages={116963},
  year={2026},
  publisher={Elsevier}
}

@article{speagle2020dynesty,
  title={dynesty: a dynamic nested sampling package for estimating Bayesian posteriors and evidences},
  author={Speagle, Joshua S},
  journal={Monthly Notices of the Royal Astronomical Society},
  volume={493},
  number={3},
  pages={3132--3158},
  year={2020},
  publisher={Oxford University Press}
}

@article{Adolfsson1996,
   author = {Lars G. Adolfsson and Bo \r{A}.S. Gustafson and Carl D. Murray},
   doi = {10.1006/icar.1996.0007},
   issn = {00191035},
   issue = {1},
   journal = {Icarus},
   month = {1},
   pages = {144-152},
   title = {The Martian Atmosphere as a Meteoroid Detector},
   volume = {119},
   year = {1996},
}

@article{Wheeler2018,
    title = {Atmospheric energy deposition modeling and inference for varied meteoroid structures},
    journal = {Icarus},
    volume = {315},
    pages = {79-91},
    year = {2018},
    issn = {0019-1035},
    doi = {https://doi.org/10.1016/j.icarus.2018.06.014},
    url = {https://www.sciencedirect.com/science/article/pii/S0019103518301313},
    author = {Lorien F. Wheeler and Donovan L. Mathias and Edward Stokan and Peter G. Brown},
}

@article{Li2026,
    author = {Li, Ziwen and Gan, Qingbo and Zeng, Xiangyuan},
    title = {Atmospheric disintegration of irregular-shaped meteoroids with sintered discrete element model},
    journal = {Meteoritics \& Planetary Science},
    volume = {61},
    number = {5},
    pages = {849-873},
    doi = {https://doi.org/10.1111/maps.70143},
    url = {https://onlinelibrary.wiley.com/doi/abs/10.1111/maps.70143},
    eprint = {https://onlinelibrary.wiley.com/doi/pdf/10.1111/maps.70143},
    year = {2026},
}

@article{popova2019modelling,
  title={Modelling the entry of meteoroids},
  author={Popova, Olga and Borovicka, Jir{\'\i} and Campbell-Brown, M},
  journal={Meteoroids: Sources of Meteors on Earth and Beyond},
  volume={25},
  number={9},
  year={2019},
  publisher={Cambridge University Press}
}

@article{Hartwick2019_noctudileMars,
   author = {V. L. Hartwick and O. B. Toon and N. G. Heavens},
   doi = {10.1038/s41561-019-0379-6},
   issn = {17520908},
   issue = {7},
   journal = {Nature Geoscience},
   month = {7},
   pages = {516-521},
   publisher = {Nature Publishing Group},
   title = {High-altitude water ice cloud formation on Mars controlled by interplanetary dust particles},
   volume = {12},
   year = {2019},
}

@article{Flynn1990mars,
   author = {G. J. Flynn and D. S. McKay},
   doi = {10.1029/jb095ib09p14497},
   issn = {01480227},
   issue = {B9},
   journal = {Journal of Geophysical Research},
   title = {An assessment of the meteoritic contribution to the Martian soil},
   volume = {95},
   year = {1990},
}

@article{ceplecha1998meteorTheory,
  title={Meteor phenomena and bodies},
  author={Ceplecha, Zden{\v{e}}k and Borovi{\v{c}}ka, Ji{\v{r}}{\'I} and Elford, W Graham and ReVelle, Douglas O and Hawkes, Robert L and Porub{\v{c}}an, Vladim{\'I}r and {\v{S}}imek, Milo{\v{s}}},
  journal={Space Science Reviews},
  volume={84},
  number={3},
  pages={327--471},
  year={1998},
  publisher={Springer}
}

@book{opik1958physics,
  title={Physics of meteor flight in the atmosphere},
  author={Opik, Ernst Julius},
  year={1958},
  publisher={Courier Corporation}
}

@article{Molina2008MetallayersMARSVENUSTITAN,
   author = {J. G. Molina-Cuberos and J. J. López-Moreno and F. Arnold},
   doi = {10.1007/s11214-008-9340-5},
   issn = {00386308},
   issue = {1-4},
   journal = {Space Science Reviews},
   month = {6},
   pages = {175-191},
   title = {Meteoric layers in planetary atmospheres},
   volume = {137},
   year = {2008},
}

@article{Whalley2010, title={Meteoric ion layers in the Martian atmosphere}, volume={147}, ISSN={1359-6640}, DOI={10.1039/c003726e}, journal={Faraday Discussions}, author={Whalley, Charlotte L. and Plane, J.M.C.}, year={2010}, pages={349} }

@article{Plane_Flynn2018, title={Impacts of Cosmic Dust on Planetary Atmospheres and Surfaces}, volume={214}, ISSN={15729672}, DOI={10.1007/s11214-017-0458-1}, abstractNote={© 2017, The Author(s). Recent advances in interplanetary dust modelling provide much improved estimates of the fluxes of cosmic dust particles into planetary (and lunar) atmospheres throughout the solar system. Combining the dust particle size and velocity distributions with new chemical ablation models enables the injection rates of individual elements to be predicted as a function of location and time. This information is essential for understanding a variety of atmospheric impacts, including: the formation of layers of metal atoms and ions; meteoric smoke particles and ice cloud nucleation; perturbations to atmospheric gas-phase chemistry; and the effects of the surface deposition of micrometeorites and cosmic spherules. There is discussion of impacts on all the planets, as well as on Pluto, Triton and Titan.}, number={1}, journal={Space Science Reviews}, publisher={The Author(s)}, author={Plane, J.M.C. and Flynn, G.J. and Määttänen, Anni and Moores, John E. and Poppe, A. R. and Carrillo-Sanchez, J. D. and Listowski, Constantino}, year={2018}, pages={1–42} }

@article{Domokos2007,
    title = {{Measurement of the meteoroid flux at Mars}},
    year = {2007},
    journal = {Icarus},
    author = {Domokos, A. and Bell, J and Brown, P G and Lemmon, M and Suggs, R. M. and Vaubaillon, J. and Cooke, W.J.},
    month = {11},
    pages = {141--150},
    volume = {191},
    url = {http://linkinghub.elsevier.com/retrieve/pii/S0019103507001923},
    doi = {10.1016/j.icarus.2007.04.017},
    issn = {00191035}
}

@incollection{Christou_Vaubaillon2019,
  author    = {Christou, A. A. and Vaubaillon, J. and Withers, Paul and Hueso, R. and Killen, Rosemary M.},
  title     = {Extra-Terrestrial Meteors},
  booktitle = {Meteoroids: Sources of Meteors on Earth and Beyond},
  editor    = {Asher, David J. and Ryabova, Galina O. and Campbell-Brown, Margaret D.},
  publisher = {Cambridge University Press},
  address   = {Cambridge},
  year      = {2019},
  pages     = {119--135},
  series    = {Cambridge Planetary Science},
  isbn      = {978-1-108-42671-8},
  doi       = {10.1017/9781108606462.011},
  url       = {https://www.cambridge.org/core/books/meteoroids/extraterrestrial-meteors/C4AA1022B3CFF1E39B1B2D9C2A2D5E77}
}

@article{henych2026geminids,
  title={Geminids are initially cracked by atmospheric thermal stress},
  author={Henych, Tom{\'a}{\v{s}} and Borovi{\v{c}}ka, Ji{\v{r}}{\'\i} and {\v{C}}apek, David and Voj{\'a}{\v{c}}ek, Vlastimil and Spurn{\`y}, Pavel and Koten, Pavel and Shrben{\`y}, Luk{\'a}{\v{s}}},
  journal={Astronomy \& Astrophysics},
  volume={706},
  pages={A26},
  year={2026},
  publisher={EDP Sciences}
}



\end{document}